\documentstyle[twoside,fleqn,espcrc2]{article}

\newcommand{\nt}{\ifmmode N_\tau\else$N_\tau$\fi}
\newcommand{\tc}{\ifmmode T_c\else$T_c$\fi}
\newcommand{\GAmma}{{\scriptscriptstyle\Gamma}}
\newcommand{\ppbar}
{\ifmmode\langle\bar\psi\psi\rangle\else$\langle\bar\psi\psi\rangle$\fi}
\newlength{\digitwidth} 
\catcode`?=\active \def?{\kern\digitwidth}

\newcommand{\AmS}{{\protect\the\textfont2
  A\kern-.1667em\lower.5ex\hbox{M}\kern-.125emS}}

\title{The Degrees of Freedom in Hot Quenched QCD}

\author{Sourendu Gupta\address{HLRZ, c/o KFA J\"ulich,
   D-5170 J\"ulich, Germany }}

\begin{document}

\begin{abstract}
In quenched QCD, on lattices with $\nt=4$, the absence of a pole in the
pseudoscalar meson channel for $T>T_c$ is demonstrated. A set of effective
4-fermi couplings is extracted. It is observed that this coupling is
small in the vector channel for all $T>T_c$, but not in the pseudoscalar
channel. The temperature dependence of hadronic parameters below \tc{} is
found to be small.\\
(Talk presented at the Lattice '92 conference, Amsterdam, September 1992.)
\end{abstract}

\maketitle

\section{Introduction}

{\begin{table*}[t]
\setlength{\tabcolsep}{1.5pc}
\caption{Screening masses in the chiral limit. A free fermion theory give
   a mass $1.32$ for $N_\tau=4$.}
\label{tab:scrm}
\begin{tabular}{rrrrr}
\hline
  $\beta$ & $\mu_\pi a$ & $\mu_\sigma a$ & $\mu_\rho a$ & $\mu_A a$ \\
\hline
$5.5???$ & $0.005\pm0.002$ &                 &
           $1.47\pm0.02$ &               \\
$5.75??$ & $0.786\pm0.002$ & $0.847\pm0.007$ &
           $1.34\pm0.01$ & $1.32\pm0.05$ \\
$5.8941$ & $1.05?\pm0.01?$ & $1.08?\pm0.01?$ &
           $1.33\pm0.03$ & $1.37\pm0.05$ \\
$6.05??$ & $1.13?\pm0.01?$ & $1.14?\pm0.02?$ &
           $1.36\pm0.02$ & $1.38\pm0.04$ \\
\hline
\end{tabular}
\end{table*}}

We present computations of screening masses and couplings both above and
below the finite-temperature phase transition in quenched QCD with $\nt=4$.
We investigate the $\rho$ and the pion below \tc, extracting the meson
masses as well as the pion decay constant $f_\pi$. We show that data
indicate the absence of mesons above \tc. We extract effective couplings
above \tc, and indicate how one can obtain lattice estimates for dimuon
cross sections in the plasma phase.

We lay special emphasis on the measurement of mesonic susceptibilities
\begin{equation}\chi_\GAmma\;=\;G_\GAmma(0).\end{equation}
Here $\Gamma$ denotes spin-flavour quantum numbers, and $G(0)$ is the
momentum-space correlator at zero four-momentum. If the low-lying
excitations in the channel $\Gamma$ are bosonic, then
\begin{equation}\chi_\GAmma\;\sim\;m_\GAmma^{-\gamma/\nu},\end{equation}
where $m_\GAmma$ is the mass of the meson and the exponent is the usual
ratio of anomalous dimensions. In free field theory this exponent has
value 2. Recall that, even in a purely bosonic theory, interactions give
rise to a cut in addition to the pole (via loop corrections to the
2-point function) which is instrumental in changing $\gamma/\nu$ from its
free field value. However, when the theory admits no pole in the channel
$\Gamma$, the spectral representation of $g_\GAmma$ has only cuts, and
$\chi_\GAmma$ goes to a constant as the screening mass is extrapolated
to zero.

In order to construct the susceptibilities, we work with correlators in
which spins and chiralities are projected out. These are constructed from
the usual $S$, $PS$, $VT$ and $PV$ correlators by the methods of
\cite{schierholz}, and denoted by the $SU(2)$-flavour notation $\pi$,
$\sigma$, $\rho$ and $A$ (although we use 4-flavour staggered fermions).

The analysis of susceptibilities involves a variation of the screening mass
(in lattice units) at fixed temperature. In the pion channel, this requires
simply a variation of the quark mass at fixed $\beta$. In both the low and
high temperature phases of the theory, a wide range of $\mu_\pi$ can be
reached by this means. For a similiar variation of $\mu_\rho$, it is not
sufficient to tune the quark mass; a variation of $\beta$ is required. Thus
a study of the $\rho$ channel similiar to the analysis we present here for
pions requires a more ambitious computation on a sequence of lattices with
changing $N_\tau$ and $\beta$ but fixed $T$. We have not attempted to do
this.

Our runs have been performed on lattices with $N_\tau=4$, and with one of
the spatial sizes large ($16\le N_z\le32$). Thus we have been able to
follow correlations to distances upto $4/T$. This is necessary if one
wants to see possible non-perturbative effects at high temperatures.
We have constructed meson screening correlators for several different
quark masses ($0.01\le m_q\le0.15$) in order to perform extrapolations
to the chiral limit. In the low-temperature phase the square of the pion
mass extrapolates to zero against $m_q$, whereas in all other cases
the screening mass extrapolates linearly against $m_q$. Full details of
the runs are available in \cite{me}.

\section{Results in the low-temperature phase}

Results below \tc{} come from measurements on $4\times8^2\times16$ and
$4\times8^2\times32$ lattices at $\beta=5.5$ (corresponding to $T=0.75T_c$).
There is little volume dependence in our results. Estimates of the screening
masses in the zero quark-mass limit are given in Table 1. The $\sigma$ and
$A$ masses are difficult to extract in this phase. The value of $m_\rho$
we obtain at $T=0.75T_c$ is in agreement with the $T=0$ results of
\cite{fuku}. We find a massless pion, indicating broken chiral symmetry.
This is confirmed by the $m_q=0$ extrapolation
\begin{equation}\ppbar a^3\;=\;\cases{
       0.227\pm0.006  & ($N_z=16$)\cr
       0.231\pm0.005  & ($N_z=32$).}\end{equation}

{}From fits of the form $\mu_\pi^2a^2=A_\pi m_qa+B_\pi$, we find
\begin{equation}A_\pi\;=\;\cases{
         6.6\pm0.2   & ($N_z=16$)\cr
         6.60\pm0.04 & ($N_z=32$),}\end{equation}
in complete consistency with the $T=0$ measurements of \cite{fuku}.
Assuming that the Gell-Mann-Oakes-Renner identity holds also at finite
temperature, we can extract the value of $f_\pi$ using the measured
value of \ppbar{} and the slope $A_\pi$ obtained from the fits above.
We find
\begin{equation}f_\pi a\;=\;\cases{
      0.185\pm0.007 & ($N_z=16$)\cr
      0.187\pm0.002 & ($N_z=32$).}\end{equation}
This is smaller than the corresponding zero temperature measurement at the
same coupling. The temperature dependence of $f_\pi$ comes entirely from
that of $\ppbar$, since $A_\pi$ is temperature independent.

We find a physical
pion state giving rise to a pole in the spectral function. On the lattice
this manifests itself as a $1/\mu\sinh\mu$ dependence of $\chi$ on the
measured screening mass. In fact, a fit of the form
$1/2\chi=A\mu\sinh\mu+B$ yields
\begin{equation}B=\cases{
    0.0008\pm0.0004 & \qquad($N_z=16$)\cr
    0.0008\pm0.0003 & \qquad($N_z=32$).}\end{equation}
Finite-volume effects are invisible, being entirely within the error bars.

{\begin{table}[b]
\setlength{\tabcolsep}{0.3pc}
\caption{Effective 4-fermi couplings.}
\label{tab:coup}
\begin{tabular}{rrrr}
\hline
    Lattice & $\beta$ & $g_{\gamma_5} T^2$ & $g_{\gamma_\mu} T^2$ \\
\hline

 $?4\times8^2\times16$ & $5.75??$ & $1.19\pm0.03$ & $?0.21\pm0.03$ \\
                       & $5.8941$ & $0.61\pm0.04$ & $?0.03\pm0.04$ \\
                       & $6.05??$ & $0.38\pm0.02$ & $-0.02\pm0.03$ \\

 $4\times12^2\times24$ & $5.75??$ & $1.10\pm0.05$ & $?0.14\pm0.04$ \\
                       & $6.05??$ & $0.45\pm0.02$ & $?0.03\pm0.03$ \\

 $4\times16^2\times32$ & $5.75??$ & $1.19\pm0.06$ & $?0.21\pm0.05$ \\
                       & $6.05??$ & $0.39\pm0.03$ & $-0.03\pm0.03$ \\
\hline
\end{tabular}
\end{table}}

\section{Results in the high-temperature phase}

These results come from simulations on $4\times8^2\times16$,
$4\times12^2\times24$ and $4\times16^2\times32$ lattices at $\beta=5.75$,
$5.8941$ and $6.05$ (corresponding respectively to $1.1$, $1.5$ and
2\tc). Our measurements of the screening masses (shown in Table 1)
at high temperatures are consistent with earlier measurements in the
quenched theory \cite{qch}. The $\rho$ and $A$ screening masses are in
good agreement with free-fermion results \cite{us}. In fact, this
agreement extends beyond the screening mass to the full correlation
function.

Susceptibility measurements strongly indicate the absence of a pole in the
pion spectral density above $T_c$. At $\beta=5.75$, a fit of the form
$1/2\chi=A\mu\sinh\mu+B$ yields
\begin{equation}B=\cases{
    0.041\pm0.003 &($4\times12^2\times24$ lattice)\cr
    0.043\pm0.004 &($4\times16^2\times32$ lattice).}\end{equation}
Thus, a pion pole is ruled out to $10\sigma$. Finite-size
effects are strong on our smallest lattice, but not on the two larger
volumes quoted above.

In the absence of bosonic poles, the chiral symmetry of the theory must
be represented in terms of an effective fermionic theory. The general
form of such an action is
\begin{equation}S_{eff}\;=\; \int_x \bar\psi D\psi +
          \sum g_\GAmma (\bar\psi\Gamma\psi)^2,\end{equation}
where $D$ is the Dirac operator, $g_\GAmma$ are effective 4-fermi
couplings, and $\Gamma$ are spin-flavour projectors. We define the
effective couplings by summing ladder graphs for $\chi_\GAmma$, and
inverting the relation to obtain $g_\GAmma$. The results of these
measurements are listed in Table 2. Parity related couplings are
equal. It is
an interesting cross-check that, in the context of such 4-fermi theories,
the derived values of the couplings require a chirally symmetric vacuum.
This effective action thus summarises the information contained in the
measurement of hadronic correlators for $T>\tc$. It remains to be checked
whether the `spatial wavefunctions' reported in \cite{them} can be
accomodated in this picture.

Just as a $SU(N_f)\times SU(N_f)$ chiral model provides an effective theory
for low energy QCD for $T<\tc$, the 4-fermi theory with the couplings given
here can be used to examine low-energy processes at $T>\tc$. In the absence
of mesonic poles in correlations of currents bilinear in the quark field, it
becomes possible to measure many interesting rate processes on the lattice.
An example is provided by the rate for low-mass dimuon production. This is
obtained from a vector correlator. This spatial correlation function can be
measured on the lattice and Wick-rotated using the information
obtained here on the spectral density function.
%

\end{document}